\documentclass[aps,prd,twocolumn,floatfix,nofootinbib]{revtex4}   

\usepackage{amsmath}    
\usepackage{graphicx}   

\newcommand{\ket}[1]{|#1\rangle}
\newcommand{\braket}[2]{\langle #1|#2\rangle}
\newcommand{\opbraket}[3]{\langle #1|#2|#3\rangle}
\begin{document}

\title{The Schroedinger equation as the time evolution \\ of a deterministic and reversible process}
\author{Gabriele Carcassi}
 \affiliation{Brookhaven National Laboratory, Upton, NY 11973}
 \email{carcassi@bnl.gov}
\date{September 4, 2008}

\begin{abstract}
In this paper we derive the Schroedinger equation by assuming it describes the time evolution of a deterministic and reversible process that leaves at each moment in time a different observable well defined; that is, it allows an accurate prediction for it. We also derive an expression for the well defined observable and apply this expression to the cases of spin precession and a free particle.
\end{abstract}

\maketitle

\section{Introduction}

In quantum mechanics textbooks, the Schroedinger equation is usually introduced by either postulating it \cite{post1,post2,post3,post4,post5,post6}, possibly accompanied by heuristic considerations \cite{heur1,heur2,heur3}, or by deriving it using mathematical arguments \cite{sakurai}. In this paper we present an alternative way that we believe is appropriate in the context of an undergraduate or graduate course. The main advantage is that it starts from broad physical assumptions, determinism and reversibility, and relates time evolution to the projection postulate, which describes the state change during a measurement.

Let us recall what is meant by determinism in the context of the Schroedinger equation, given that quantum mechanics is, as a whole, non-deterministic. From an initial state, the Schroedinger equation predicts what state the system will be in at each particular moment in time. This state is predicted exactly. Observables, instead, are not exactly predicted: the state in general will define a probability distribution for each. As the state is deterministically predicted, the distributions are as well, yet not the actual values.

\section{Infinitesimal time evolution}

Before we start looking for an expression that describes the state evolution in time, we note that any state is an eigenstate of at least one Hermitian operator. We will demonstrate this by showing that if there exists at least one state that is an eigenstate of a Hermitian operator, then all states are eigenstates of some Hermitian operator.

Let $\ket{\Psi_a}$ and $\ket{\Psi_b}$ be two normalized vectors in a Hilbert space $\mathcal{H}$, and let the first be an eigenket of Hermitian operator $A$. Given that they are normalized, we can always define a unitary operator $U$ such that $\ket{\Psi_b}=U\ket{\Psi_a}$: we do so by rotating on the plane defined by the two vectors and leaving the other directions unchanged. If we now consider the operator $B=UAU^{\dagger}$, it is easy to show that $\ket{\Psi_b}$ is an eigenket of $B$ and that $B$ is Hermitian. Given that every state is an eigenstate of a Hermitian operator, for each state there is going to be a quantity, an observable, whose value is defined with complete certainty. Regardless of what physical process we study, \emph{time evolution will leave the system with a well defined quantity at each moment in time}.

Let us now consider the state at two different but close instants in time, $\ket{\Psi, t}$ and $\ket{\Psi, t+dt}$. The system transitions from its initial state to a final state, which is an eigenstate of a Hermitian operator, and given that $dt$ is short we disregard intermediate states within the interval. We then need to characterize a process that makes the system transition to an eigenstate of a Hermitian operator, without requiring description of anything beyond the initial and the final states. We note that we already have a way to describe a process with such characteristics: the projection postulate used to describe the change of state during measurements. This says that a system will transition from the initial state $\ket{\Psi}$ to one of the eigenstates $\ket{\Psi_i}$ of the operator associated with the measured observable, and that the probability to end in each eigenstate is given by $\braket{\Psi}{\Psi_i}\braket{\Psi_i}{\Psi}$. We assume, then, that the projection postulate can also be used to describe an infinitesimal time evolution.\footnote{Note that no measurement is performed, in the usual sense of the word. Time evolution leaves these quantities defined whether they are actively measured or not. The assumption here is that the mechanism that leaves these quantities well defined during time evolution and the one that leaves them well defined during a measurement are either the same, or produce results similar enough that we can describe them in the same manner.}

Time evolution is then a series of infinitesimal steps that leave a particular quantity well determined, and the change effected by each infinitesimal step can be described by the projection postulate.

\section{Deterministic and reversible processes}

In general time evolution as defined above will be non-deterministic since the projection postulate is non-deterministic. Now we want to restrict ourselves to study processes that are deterministic, for which the final state can be determined, predicted, from the initial state. Under which conditions is this the case?

Let us assume that $A^{(t)}$ and $A^{(t+dt)}$ are two observables that are well defined at time $t$ and $t+dt$ respectively. If they are the same, the state is not going to change. More generally: if they are compatible, that is if their commutator is zero, the state is not going to change because an eigenstate of one is an eigenstate of the other. This evolution is indeed deterministic, but has no consequences. On the other hand, if the two are incompatible the state will change. The more they are incompatible, that is, the more their eigenstates differ, the greater and less predictable the change. What we need for a deterministic process is for those two observables to be almost compatible, so that the eigenstates are very close to each other, but not completely compatible. If those observables represented spin measurements, for example, they should then represent two different directions but very very close to each other.

For the process to be deterministic the eigenkets of $A^{(t+dt)}$ need to be in almost the same direction as $A^{(t)}$, except for an infinitesimal difference. Given that the initial state and the final state are eigenkets of those two operators, the inner product between them is:

\begin{equation}
\label{determinism}
\braket{\Psi, t+dt}{\Psi, t}
 = 1 - \epsilon dt
\end{equation}

Note that we started from requiring determinism and we ended up requiring continuity: a small change in time has to cause a small change in the state. Continuous evolution and determinism in this context are one and the same. The less continuous the state is in time, the more the evolution is non-deterministic.

We also require that the process be reversible. This means that if we start from the final state and go back to the initial state, the change should be equal but opposite. We have:

\begin{equation}
\label{reversibility}
\braket{\Psi, t}{\Psi, t+dt}
 = 1 + \epsilon dt
\end{equation}

These are the two expressions from which we derive the Schroedinger equation.

The two equations describe quantities that are complex conjugates of one another. We have:
\begin{align*}
1 + \epsilon dt
&= (1 - \epsilon dt)^*\\
\epsilon &= - \epsilon^*
\end{align*}
$\epsilon$ is therefore imaginary. We also note that $\epsilon$ cannot be a constant but has to depend on the state. If it did not, we could always write:
\begin{equation*}
\ket{\Psi, t+dt}
= (1 - \epsilon dt)^* \ket{\Psi, t}
\end{equation*}
 The initial state and the final state would always differ by a multiplicative constant, which means that they would always describe the same physical state: no evolution. We then write the final state as a function of the initial state, where $\Omega$ is a linear operator that describes the state change per unit of time

\begin{equation}
\label{schrprototype}
\ket{\Psi, t+dt} = 1 - i\Omega dt \ket{\Psi, t}
\end{equation}

If we combine with equation \eqref{determinism}, we see that
\begin{equation*}
\epsilon = - i \opbraket{\Psi, t}{\Omega}{\Psi, t}
\end{equation*}
which guarantees that $\epsilon$ depends on the state. It also tells us that $\Omega$ is Hermitian, because the expectation is real. With this in mind and ignoring second-order terms, we have:
\begin{align*}
&\braket{\Psi, t+dt}{\Psi, t+dt} \\
&= \opbraket{\Psi, t}{(1 - i\Omega dt)^\dagger | (1 - i\Omega dt)}{\Psi, t} \\
&= \opbraket{\Psi, t}{(1 + i\Omega^\dagger dt - i\Omega dt + \Omega^\dagger \Omega dt^2)}{\Psi, t} \\
&= \braket{\Psi, t}{\Psi, t}
\end{align*}
which tells us that a reversible process is one that conserves probability.

We can say the following about $\Omega$:
\begin{itemize}
  \item it represents an observable, as it is Hermitian
  \item it has to be a function of the observables that define the state of the system (such as position, momentum, spin, ...): $\Omega$ tells us how the state is going to change in time and, since the process is deterministic, needs to be a function of the state
  \item the expectation of $\Omega$ does not change in time, if $\Omega$ does not depend on time:
\end{itemize}
\begin{align*}
&\opbraket{\Psi, t+dt}{\Omega}{\Psi, t+dt} \\
&= \opbraket{\Psi, t}{(1 - i\Omega dt)^\dagger | \Omega | 1 - i\Omega dt}{\Psi, t} \\
&= \opbraket{\Psi, t}{\Omega | (1 - i\Omega dt)^\dagger | 1 - i\Omega dt}{\Psi, t} \\
&= \opbraket{\Psi, t}{\Omega}{\Psi, t}
\end{align*}
Empirically we find that:
\begin{equation*}
\Omega = H / \hbar
\end{equation*}
where $H$ is the Hamiltonian and $\hbar$ is the well known constant (this is imposed and the only justification is that it works)\footnote{Alternatively, one can use the arguments found in Sakurai \cite{sakurai}.}. We substitute in \eqref{schrprototype} and we have:

\begin{align}
\ket{\Psi, t+dt} = \ket{\Psi, t} &- \frac{i H dt}{\hbar} \ket{\Psi, t} \notag \\
\ket{\Psi, t+dt} - \ket{\Psi, t} &= - \frac{i H dt}{\hbar} \ket{\Psi, t} \notag \\
i \hbar \frac{\partial}{\partial t} \ket{\Psi, t} &= H \ket{\Psi, t}
\end{align}

This tells us that the Schroedinger equation can be used to describe the time evolution of a deterministic process, but this is hardly a surprise. The derivation itself is essentially equivalent to the one found in Sakurai \cite{sakurai}, except that we start from physical considerations, determinism and reversibility, instead of mathematical considerations, continuity and probability conservation.\footnote{Sakurai requires continuity but does not make the following, in our opinion important, distinction. When in classical mechanics we require $x(t)$ to be continuous, we mean that the \emph{value} of $x$ varies little in time: it changes to one that is closer first. Continuity in quantum mechanics is radically different: all eigenkets of the position operator are orthogonal to each other and they are all equally "distant", so requiring the evolution to be continuous does not mean that the value of a particular observable evolves continuously. What does evolve continuously is the probability distribution. This is particularly evident for spin 1/2: a small change in state will change the spin angle, which corresponds to a different observable. The spin value, instead, can never change continuously as spin is a discrete quantity.}

One cannot help but wonder: what quantities does time evolution leave determined at each moment in time?

\section{Observables in time}

As there are an infinite number of observables that share the same final states, it is not possible to univocally say what quantity is left defined by time evolution at each point in time. $x$, $x^3$ and $2x^3$ are all equivalent, because measuring one means automatically knowing all the others. This, though, means we have some room to conveniently choose a series of observables that are easy to study.

First of all, we assume that we have a measurement $A^{(t=0)}$ performed at $t=0$ giving the value $\alpha$:

\begin{equation}
A^{(t=0)}\ket{\Psi, t=0} = \alpha \ket{\Psi, t=0}
\end{equation}

Now we move to another moment in time. The system will be in the eigenstate of an infinite number of operators corresponding to all observables that are well defined at that particular instant. For convenience we will choose the one for which the eigenvalues are the same as for $A^{(t=0)}$. That is, the observable changed but the actual quantity did not. We have:

\begin{equation}
A^{(t)}\ket{\Psi, t} = \alpha \ket{\Psi, t}
\end{equation}

Using the solution to the Schroedinger equation, we have:
\begin{equation*}
A^{(t)} e^{-i H t / \hbar} \ket{\Psi, t=0} = \alpha e^{-i H t / \hbar} \ket{\Psi, t=0}
\end{equation*}
We multiply both sides by $e^{i H t / \hbar}$ and have:
\begin{align*}
e^{i H t / \hbar} &A^{(t)} e^{-i H t / \hbar} \ket{\Psi, t=0} \\
&= e^{i H t / \hbar} \alpha e^{-i H t / \hbar} \ket{\Psi, t=0} \\
  &= \alpha \ket{\Psi, t=0} \\
  &= A^{(t=0)} \ket{\Psi, t=0}
\end{align*}
which means:
\begin{equation}
\label{obstime}
A^{(t)} = e^{-i H t / \hbar} A^{(t=0)} e^{i H t / \hbar}
\end{equation}
We note that at $t=0$ $A^{(t)}$ corresponds to $A^{(t=0)}$ as one would expect.

Let us not get confused: we are still in the Schroedinger picture, where the \emph{state} changes in time. $A^{(t)}$ describes a different operator at each instant in time, but also describes a different \emph{observable} at each instant in time (unlike what is normally done in the Heisenberg picture). It describes a series of observables that are going to be fully determined during the evolution, which one would be able to predict with complete certainty. The value of those will always be the same because we chose the operator with that property.

\section{Spin precession}

To better illustrate the ideas expressed in the last section, we study a well known case: spin precession \cite{sakuraispin}. An electron is in a constant magnetic field aligned with the z axis. If the initial spin is aligned with the z axis, it will not change; if it is aligned with the x axis, it will rotate in the x-y plane.

Intuitively, we can see how the picture we developed before fits. If the spin of the electron is initially aligned with the z axis, each time step leaves the spin well defined along that direction. If the spin is initially along the x axis, each time step leaves the spin defined a little bit further counterclockwise, thus changing the direction at each moment in time, making the spin rotate.

Let us now go through the equation and see how it works. The Hamiltonian for the system is:
\begin{align}
H &= - \frac{e}{m_e c} \mathbf{S}\cdot\mathbf{B} \notag \\
&= - \frac{e B_z}{m_e c} S_z \notag \\
&= \omega S_z
\end{align}
where
\begin{equation*}
\omega = \frac{|e| B_z}{m_e c}
\end{equation*}
We substitute the Hamiltonian in expression \eqref{obstime}, and we have:
\begin{align}
A^{(t)} &= e^{-i \omega S_z t / \hbar} A^{(t=0)} e^{i \omega S t / \hbar} \notag \\
&= e^{\frac{-i \omega t}{\hbar} \frac{\hbar}{2}
\begin{bmatrix}
1 & 0 \\
0 & -1 \hfill
\end{bmatrix}
} A^{(t=0)}
e^{\frac{-i \omega t}{\hbar} \frac{\hbar}{2}
\begin{bmatrix}
1 & 0 \\
0 & -1 \hfill
\end{bmatrix}
} \notag \\
&=
\begin{bmatrix}
e^{\frac{-i \omega t}{2}} & 0 \\
0 & e^{\frac{i \omega t}{2}} \hfill
\end{bmatrix}
A^{(t=0)}
\begin{bmatrix}
e^{\frac{i \omega t}{2}} & 0 \\
0 & e^{\frac{-i \omega t}{2}} \hfill
\end{bmatrix}
\end{align}

Now if we assume that we start by having measured the spin in the z direction, we substitute $A^{(t=0)}$ with $S_z$. We obtain:
\begin{align}
A^{(t)} &=
\begin{bmatrix}
e^{\frac{-i \omega t}{2}} & 0 \\
0 & e^{\frac{i \omega t}{2}} \hfill
\end{bmatrix}
\frac{\hbar}{2}
\begin{bmatrix}
1 & 0 \\
0 & -1 \hfill
\end{bmatrix}
\begin{bmatrix}
e^{\frac{i \omega t}{2}} & 0 \\
0 & e^{\frac{-i \omega t}{2}} \hfill
\end{bmatrix} \notag \\
&=
\frac{\hbar}{2}
\begin{bmatrix}
1 & 0 \\
0 & -1 \hfill
\end{bmatrix} \notag \\
&= S_z
\end{align}

As expected, the time evolution will simply keep spin z defined at each moment in time, which will make the state remain the same. If we assume that we start by having measured the spin in the x direction we obtain:
\begin{align}
A^{(t)} &=
\begin{bmatrix}
e^{\frac{-i \omega t}{2}} & 0 \\
0 & e^{\frac{i \omega t}{2}} \hfill
\end{bmatrix}
\frac{\hbar}{2}
\begin{bmatrix}
0 & 1 \\
1 & 0 \hfill
\end{bmatrix}
\begin{bmatrix}
e^{\frac{i \omega t}{2}} & 0 \\
0 & e^{\frac{-i \omega t}{2}} \hfill
\end{bmatrix} \notag \\
&=
\frac{\hbar}{2}
\begin{bmatrix}
0 & e^{-i \omega t} \\
e^{i \omega t} & 0 \hfill
\end{bmatrix} \notag \\
&=
\frac{\hbar}{2}
\begin{bmatrix}
0 & \cos(- \omega t) + i \sin(- \omega t) \\
\cos(\omega t) + i \sin(\omega t) & 0 \hfill
\end{bmatrix} \notag \\
&=
\cos(\omega t) S_x + \sin(\omega t) S_y
\end{align}
Here we can clearly see the precession, as the direction left defined will keep rotating in the x-y plane.

\section{Free particle}

Another instructive example is the case of the free particle. Here the Hamiltonian is given by:
\begin{equation*}
H = \frac{P^2}{2 m}
\end{equation*}
We start in a state where the position is determined, and remembering that:
\begin{equation*}
[F(P), X] = - i \hbar \frac{\partial F(P)}{\partial P}
\end{equation*}
we have:
\begin{align}
A^{(t)} &= e^{-i H t / \hbar} X e^{i H t / \hbar} \notag \\
&= X + [ e^{-i H t / \hbar}, X] e^{i H t / \hbar} \notag \\
&= X -i \hbar (-i \frac{2Pt}{2m\hbar}) \notag \\
&= X - \frac{Pt}{m}
\end{align}

This means that if at time $t$ we measure the quantity $X-Pt/m$, we will have a precise prediction which will be equal to the initial position $x_0$. We can write:
\begin{align*}
(X - \frac{Pt}{m})\ket{\Psi, t} = x_0 \ket{\Psi, t} \\
X \ket{\Psi, t} = (x_0 + \frac{Pt}{m})\ket{\Psi, t}
\end{align*}
which, though it looks similar to a free particle trajectory in classical mechanics, is a relationship between the operators so the physical meaning is very different. If we start with the momentum determined, given that it will commute with the Hamiltonian, we simply have:
\begin{align}
P \ket{\Psi, t} = p_0 \ket{\Psi, t}
\end{align}
which means that the momentum will not change.

\section{A note on interpretations}

We want to point out that all that we have said is independent of what interpretation one may choose. Shall we assume that the wave function collapses during a measurement? Then the evolution can be seen as a series of infinitesimal collapses. Shall we assume that there are hidden variables? Then they will simply change throughout a physical process with the same rules that describe their change during a measurement. But there is also nothing in our discussion that prohibits one from interpreting time evolution and the projection postulate as describing two completely different physical processes, if one chooses to do so: the fact that mathematically one can be described as the limit of the other under certain conditions does not have to imply that they are physically related.

Regardless of which interpretation one may choose, the important point is that the expression for the fully defined observables is an actual prediction, and as such has to be valid for any interpretation of quantum mechanics.

\section{A simplified picture}

We believe that this picture may allow us to conceptually describe the evolution of a quantum system in a way that can be simplified but still be made rigorous. Once one accepts the fact that there are quantities that are not compatible, that if we measure one (e.g. position) some other (e.g. momentum) is unknown, we can concentrate only on the quantity that can be known at each moment of time. If we do this, we may get away with not having to talk about complex vector spaces and inner products.

For example, let us say we start at $t=0$ with $x=x_0$ and therefore $p$ undefined. After some time $t=\Delta t$ there is going to be a quantity that is left defined, $x - p \Delta t/m$, that can be fully determined by knowing $x_0$. Any other quantity that could be used to determine both $x$ and $p$ (such as $x$, $p$, $x+p$, ...) cannot be known exactly. If we have an observable $f(x,p,t)$ that is fully known, we will also have an incompatible observable $g(x,p,t)$ completely unknown. In these cases, quantum uncertainty is described by the fact that we are missing one equation, which can be understood by simply requiring mathematical understanding of systems of algebraic equations in two unknowns. We also note that we cannot really see the degree to which two observables are incompatible, which is what we lose with this simplification. Nevertheless, we can see in the following cases how some of the behavior characteristic of quantum mechanics has to emerge.

If we start with $p=p_0$, the final quantity $f(x,p,t)$ is not going to depend on an initial position as it will be deterministically predicted by the initial value of momentum, regardless of what the actual physical process (the Hamiltonian) is. This shows already how non-locality plays an important role in quantum mechanics.

If we describe the state of two independent systems, we will now have 2 equations in 4 unknowns:
\begin{align*}
f_1(x_1,p_1,t) &= b_1 \\
f_2(x_2,p_2,t) &= b_2
\end{align*}
If the systems are instead not independent, we will need to have
\begin{align*}
f_1(x_1,p_1,x_2,p_2,t) &= b_1 \\
f_2(x_1,p_1,x_2,p_2,t) &= b_2
\end{align*}
which tells us that position and momentum of the two systems will somehow influence each other. But it also tells us that the evolution of the two systems cannot always be described as the two systems separately. For example, if our initial observables are $x_1-x_2$ and $p_1+p_2$, the separate position and momentum of each system is unknown (having any one would allow us to have 3 equations in 4 unknowns), and therefore the evolution of the overall system cannot depend on it: the systems are entangled.

Quite naturally more work needs to be done along this path as this is just a sketch. And it should be clear that not all interesting cases will be able to fit into this simplified picture. But if enough examples are found, it may be a tool to introduce some aspects of quantum mechanics assuming only a mathematical background of systems of equations. This could be considered analogous to the way in which some aspects of classical mechanics are introduced at the high school level without calculus. We look forward to suggestions and guidance from the physics education community.

\section{Conclusion}

We have seen that the Schroedinger equation can be derived by assuming that we are describing the time evolution of a deterministic and reversible process comprised of a series of infinitesimal time steps, each leaving a quantity well defined. This derivation gives, at least in our opinion, a more satisfying picture because it starts from very broad physical assumptions, and it shows how the changes of state described by the Schroedinger equation and the projection postulate can be related. We hope that others will find this approach to be an instructive alternative to the more standard ways of introducing the Schroedinger equation found in the current literature.

\section{Acknowledgments}

We would like to thank Christine Aidala, Boaz Nash, Elemer Elad Rosinger, Anne Sickles, Paul Stankus and George Sterman for discussions which helped review and clarify the ideas presented in this paper.

\end{document}